\documentclass[12pt,usenatbib]{mn2e}
\usepackage{graphicx}
\usepackage{epsf}
\usepackage{psfig}

\def\lsim{\mathrel{\lower0.6ex\hbox{$\buildrel {\textstyle <}
 \over {\scriptstyle \sim}$}}}
\def\gsim{\mathrel{\lower0.6ex\hbox{$\buildrel {\textstyle >}
 \over {\scriptstyle \sim}$}}}
\def\Jdm{\mathrel{\hat{\rm \underline{J}}_{\rm dm}}}
\def\Jsat{\mathrel{\hat{\rm \underline{J}}_{\rm sat}}}
\def\Jdisc{\mathrel{\hat{\rm \underline{J}}_{\rm gal}}}

\def\cdm{\mathrel{\hat{\rm \underline{c}}_{\rm dm}}}

\def\cdisc{\mathrel{\hat{\rm \underline{c}}_{\rm gal}}}

\def\csat{\mathrel{\hat{\rm \underline{c}}_{\rm sat}}}

\def\asat{\mathrel{\hat{\rm \underline{a}}_{\rm sat}}}

\begin{document}

\title[Alignment of Satellite Galaxies]{Satellite Systems around 
  Galaxies in Hydrodynamic Simulations}
\author[N. I. Libeskind et al.]{
\parbox[h]{\textwidth}{Noam I Libeskind$^{1}$,
  Shaun Cole$^{1}$,  Carlos S Frenk$^{1}$, Takashi
  Okamoto$^{1,2}$, and Adrian Jenkins$^{1}$}
\vspace{6pt} \\
$^{1}$Department of Physics, University of Durham, Science
  Laboratories, South Road, Durham, DH1 3LE, U.K.\\
$^{2}$National Astronomical Observatory of Japan, Mitaka, Tokyo
  181-8588 \\
}


\maketitle
\begin{abstract}

We investigate the properties of satellite galaxies formed in
N-body/SPH simulations of galaxy formation in the $\Lambda$CDM
cosmology. The simulations include the main physical effects thought
to be important in galaxy formation and, in several cases, produce
realistic spiral discs. In total, a sample of 9 galaxies of luminosity
comparable to the Milky Way was obtained.  At magnitudes brighter than
the resolution limit, $M_V=-12$, the luminosity function of the
satellite galaxies in the simulations is in excellent agreement with
data for the Local Group. The radial number density profile of the
model satellites, as well as their gas fractions also match
observations very well. In agreement with previous N-body studies, we
find that the satellites tend to be distributed in highly flattened
configurations whose major axis is aligned with the major axis of the
(generally triaxial) dark halo. In 2 out of 3 systems with
sufficiently large satellite populations, the satellite system is
nearly perpendicular to the plane of the galactic disc, a
configuration analogous to that observed in the Milk Way. The discs
themselves are perpendicular to the minor axis of their host halos in
the inner parts, and the correlation between the orientation of the
galaxy and the shape of the halo persists even out to the virial
radius. However, in one case the disc's minor axis ends up, at the
virial radius, perpendicular to the
minor axis of the halo. The angular momenta of the galaxies and their
host halo tend to be well aligned.  \end{abstract}


\section{Introduction}
\label{introduction}
A successful theory of galaxy formation needs to explain a wide
variety of physical phenomena across many scales. On the Galactic mass
scale, the problem of how small satellites form around galaxies such
as the Milky Way poses two interesting, and as yet not fully answered
questions. These are: \newline \indent 1) What determines the number
of Milky Way satellites and the shape of
  their luminosity function?  
\newline
\indent
2) Why are the Milky Way's satellites aligned on a great circle in
   the sky and why is this great circle nearly perfectly perpendicular to
   the disc of the Milky Way?  \newline \indent At present, it is
   unclear whether the luminosity function of the satellites of the
   Milky Way and their peculiar spatial distribution are unique to the
   Galaxy, or whether these are generic features that arise during the
   formation of a typical galaxy. By virtue of their small size,
   extragalactic satellites are difficult to observe and most studies
   are hampered by either flux limits, completeness, or sample
   size. However, recent advances using large sky surveys have
   resulted large samples -- albeit with only very few satellites per
   primary -- and have shed some light on these questions.

In a universe in which the gravitationally dominant component is 
cold dark matter (CDM), the existence of small-scale power in the
initial conditions causes the early collapse of matter; 
cosmological simulations show that small, dense, CDM haloes are able
to form at early times. These low mass haloes grow by successive
mergers and smooth accretion giving rise to the large-scale structures
we see today (e.g \citealt{Frenk85, NFW95, Wechsler02, Zhao03}). High
resolution N-body simulations of the growth of CDM 
halos have shown that the dense cores of merging clumps often survive
the disruptive effects associated with mergers and remain as distinct
substructures (or subhalos) embedded within a larger, smooth main halo
(\citealt{Klypin99}, \citealt{Moore99}). Although the total mass
attached to subhalos is only of order $\sim 10\%$ of the total halo
mass (\citealt{Ghigna98,Gao04,Springel01,Stoehr02}), both N-body
simulations and semi-analytical calculations of the assembly of dark
matter haloes based on the extended Press-Schechter theory
(\citealt{Bond91}; \citealt{Bower91}; \citealt{Lacey93}), show
that many more embedded substructures survive than there are visible
satellites in galactic halos (Kauffmann, White \& Guiderdoni
1993; \citealt{Moore99,Klypin99}).

Many authors have argued that this so-called ``missing satellite
problem'' poses a severe challenge to the CDM cosmology and encourages
the study of alternative forms of dark matter
(\citealt{Craig01,Moore00,Spergel00,Yoshida00}) or of 
different cosmological initial conditions (\citealt{Kamionkowski}). Other authors
have argued that the paucity of satellites in the Local Group is a
natural byproduct of the physics of galaxy formation that regulate the
cooling of gas in small halos (\citealt{Kauffmann93};
Bullock, Kravstov \&
Weinberg 2000; \citealt{Benson02a}) or merely the result of a
misidentification of substructures in the simulations with satellites
in the Milky Way (\citealt{Stoehr02}). The semi-analytical model of
galaxy formation of \cite{Benson02b} included a detailed treatment of
the reionization of hydrogen in the early universe which, by altering
the thermodynamic state of primordial gas, inhibits the formation of
small satellite galaxies. In this model, the increase in the entropy
of the gas has two effects: it inhibits the cooling of new gas into
small halos and it delays star formation in gas that has already
cooled. The end result is a satellite population with about the
observed numbers seen in the Local Group and with a luminosity
function that matches the faint end of the observed function but not
its bright end where the model fails to produce enough large, 
LMC-like satellites.

The spatial distribution of satellite galaxies poses another
interesting problem within the CDM paradigm. Whereas N-body
simulations show that the substructures that survive within halos tend
to be nearly spherically distributed, the galactic satellites of 
the Milky Way are confined to a highly flattened structure, a puzzling
fact first recognized 30 years ago (\citealt{LyndenBell76},
\citealt{Kunkel76}; see also \citealt{LyndenBell82}). Kroupa, Thies \&
Boily (2005) drew attention to this discrepancy and concluded that the 
anisotropic alignment of the Galaxy's satellites contradicts the CDM
model. Recent work, however, has shown that the satellite galaxies do
not populate a random selection of subhalos but are preferentially
found in a biased subset which is arranged in a flattened
configuration. This bias partly reflects the preferential infall of
the most massive dark matter clumps along the filaments of the cosmic
web. This phenomenon is clearly seen (to various degrees) in the
N-body simulations analyzed by \cite{Kang05}, \cite{Libeskind05} and
\cite{Zentner05} (and in the cluster mass simulations of
\citealt{Knebe04}).  These studies differ in the precise way in which
subhalos are identified with satellites but they all agree that
flattened satellite configurations such as that seen in the Milky Way
are not uncommon. In particular, Libeskind et al. and Zentner et
al. followed the formation of satellites by applying a semi-analytic
galaxy formation model to high-resolution N-body simulations.  Both
studies found not only fattenings consistent with that seen in the
Milky Way, but also that the long axis of the flattened satellite
distribution tends to be aligned with the long axis of the parent dark
matter halo.

Beyond the Local Group, a number of studies have claimed correlations
between the orientation of central galaxies and the distribution of
their satellites. \cite{Holmberg69} first identified a lack of
satellites in the plane of a coadded sample of central galaxies out to
a projected radius of $r_{\rm p} \lsim 50$~kpc.  \cite{Zaritsky97}
found evidence for this ``Holmberg effect'' but only on much larger
scales ($300 \lsim r_{\rm p} \lsim 500$~kpc). The reality of the
Holmberg effect remains controversial. Early authors claimed that if
an anisotropy exists at all, it is, even if significant, at best small
(\citealt{Hawley75}, \citealt{Sharp79}, \citealt{MacGillivray82}). An
enhancement perpendicular to the disc was recently inferred in the
distribution of satellites identified in the 2dFGRS by \cite{Sales04}
but a private communication quoted in \citet{Yang05} indicates that
the original analysis was incorrect and that the enhancement is, in
fact, along the disc, not perpendicular to it. \cite{Brainerd05} and
\cite{Yang05} have claimed to see an alignment of satellites in the
SDSS\footnote{http://www.sdss.org/} in the opposite direction to
Holmberg's, that is along the plane of the galaxy disc rather than
orthogonal to it.  \cite{Yang05} found that the planar distribution is
detectable only in red satellites but the blue population is
consistent with an isotropic distribution.  These apparently
conflicting observational studies all seem to suggest anisotropic
distributions of satellites but they disagree on how the overall
distribution of satellites is oriented relative to the central galaxy.

In principle, simulations are an ideal way to investigate this sort of
issues. Since their introduction to cosmology in the 1970s and early
1980s (\citealt{Peebles71}; Aarseth, Turner, \& Gott 1979, Frenk,
White \& Davis 1983), N-body simulations have been extremely useful in
revealing how cosmic structures emerge out of small primordial
perturbations (see Springel, Frenk \& White 2006 for a review).  To
investigate questions such as the alignment between satellites and
central galaxies, however, it is necessary to follow not only the
evolution of dark matter, but the coupled evolution of the baryonic
component as well. Until recently, progress in this area was hampered
by the inability of hydrodynamic simulations to produce realistic
discs from CDM initial conditions. Without some form of feedback to
prevent most of the gas from cooling into subgalactic fragments, the
outward transfer of orbital angular momentum to the dark halo as these
fragments merge results in discs that are much too small (\citealt
{NavarroBenz}; \citealt{WeilEkeEfstathiou};
\citealt{SommerLarsenGelatoVedel}; \citealt{EkeEfstathiouWright}).

The ``disc angular momentum problem'' has recently been overcome, at
least partially, in a number of simulations which include plausible
forms of feedback and are able to produce relatively realistic
galactic discs (e.g. \citealt{Abadi03}, \citealt{Governato04},
\citealt{Robertson2004}, \citealt{SommerLarsenGelatoVedel}, 
\citealt{Bailin05}, \citealt{Okamoto05}). In this paper, we analyse
the simulations carried out by \cite{Okamoto05}.  Specifically, we
investigate the properties of satellite galaxies orbiting central
galaxies of mass similar to that of the Milky Way. We derive the
satellite luminosity function over a limited, but still 
interesting, range of luminosity. We search for anisotropy in the
satellite galaxy distribution, and study the alignment of satellite
systems with their central disc, as well as the alignment of
the disc with the its host dark matter halo.

This paper is organized as follows. In Section~2 we describe
the simulations we have used, as well as our method for
selecting complete satellite samples. In that section we also derive
the satellite galaxy luminosity function and investigate the gas
fraction of the largest satellites. We present our analysis of
relative shape alignments in Section~3 and of angular momentum
alignments in Section~4, and conclude in Section~5. 

\section{Identifying galaxies and satellites}
In this section, we briefly describe the simulations that we have
analyzed and the methods that we have developed in order to identify
central and satellite galaxies.

\subsection{The simulations}
\label{simulations}

We have analyzed two simulations of galaxy formation, both carried out
using the parallel PM-TreeSPH code \texttt{GADGET2}
(\citealt{Springel05}), as modified by
\cite{Okamoto05}. \texttt{GADGET2} calculates the evolution of dark
matter using N-body techniques and the evolution of gas using smooth
particle hydrodynamics (SPH). 

Initially, the two simulations followed the evolution of dark matter
in a cosmological cubical volume of length $L_{\rm box}=35.325
h^{-1}$Mpc in a $\Lambda$CDM model with cosmological parameters
$\Lambda$=0.7, $\Omega_{\rm m}$=0.3, $H_{0}=70$~km~s$^{-1}$ and
$\sigma_{8}$=0.9. For the first simulation (hereafter SD), a region
around a suitably chosen dark matter halo was identified at the final
time and the simulation was run again, this time adding many more dark
matter particles, as well as SPH particles, in the region of the halo, and
perturbing these with additional high frequency power drawn from the
same power spectrum of the original simulation following the general
method outlined by \cite{Frenk96}. Since the goal of this simulation
was to investigate the formation of a galactic disc, the halo chosen
for resimulation was selected to have a quiet recent merger history,
with no major mergers since $z\approx 1$. A preliminary
semi-analytical calculation applied to the merger tree of this halo,
using the methods of \cite{Helly03}, indicated that a disc galaxy was
likely to form in this halo. The high resolution region enclosed a
spherical volume of radius 0.9~Mpc around the halo at $z=0$.

Simulation SD did indeed form a reasonably realistic disc, as
discussed by \cite{Okamoto05} (in their ``shock-burst''
model). Encouraged by this success, we ran a second simulation
(hereafter SR) with the same code, this time populating several
regions of the same volume with high resolution dark matter and gas
particles. The high resolution regions consisted of a large sphere of
radius 5$h^{-1}$Mpc and four smaller overlapping spheres each of
radius r=1$h^{-1}$Mpc. This arrangement ensured coverage of all the
large galaxies that formed out to the virial radii of their halos.
The same cosmological parameters were used in both simulations, except
that the baryon density was taken to be $\Omega_{\rm b}$=0.040 in SD, 
and slightly larger, $\Omega_{\rm b}$=0.044, in SR.  In both 
simulations, the mass per particle was $\sim 2.6 \times 10^{6}$ for
gas and $\sim 1.7 \times 10^{7} h^{-1}$ M$_{\odot}$ for dark matter. 
The gravitational softening was 0.5 and 1$h^{-1}$~kpc for SPH and high
resolution dark matter particles respectively, while the minimum SPH smoothing
length was 0.5 $h^{-1}$~kpc for simulation SD and 0.39 $h^{-1}$~kpc for simulation SR.

The various physical processes included in our simulations are
described in detail in \cite{Okamoto05}. Here, we summarize the
salient points. The interstellar medium (ISM) is modelled, following
\cite{SpringelHernquist03}, as a two phase medium composed of hot
ambient gas and cold gas clouds in pressure equilibrium. Gas heating
and cooling rates are computed assuming collisional ionization
equilibrium in the presence of a uniform and evolving ultra-violet
(UV) background, which is assumed to be generated by hot OB stars and
is switched on at $z=6$ (\citealt{HardtMadau}).  The cooling rates,
which depend on the metallicity of the gas, are computed from the
tables given by \cite{SutherlandDopita93}; molecular cooling and other
forms of cooling below $T \approx 10^{4}$~K are ignored.

Star formation can occur in a ``quiescent'' and a ``burst'' mode. In
the quiescent mode, gas particles that meet a specified density
criteria, are turned into stars according to a pre-determined
probability. These stars form with a standard initial mass function
(IMF; \citealt{Salpeter55}). Bursts of star formation are
triggered by major mergers which are identified by tracking large
changes in the entropy of the gas. In a burst, stars form on a shorter
timescale than in the quiescent mode and with a top-heavy IMF. This
model of star formation is motivated by the semi-analytical work of
\cite{Baugh05} who argue that only a top-heavy IMF in bursts can
explain the number density of sub-millimetre and Lyman-break galaxies
at high redshift. \cite{Nagashima05a,Nagashima05b} argue, similarly,
that this model is also required to explain the metallicity of the
intracluster medium and of elliptical galaxies.

In the \cite{Okamoto05} model, the evolution of the stellar
populations that form is followed in detail, tracking both type-II and
type-Ia supernovae. This requires abandoning the instantaneous
recycling approximation (IRA) assumed by \cite{SpringelHernquist03},
whereby star formation, cold gas cloud formation by thermal
instability, the evaporation of gas clouds and the heating of ambient
gas by supernovae explosions all occur simultaneously.  Instead,
following each star formation event, supernovae energy and
metals are injected back into the ISM on a timescale which is computed
from the mass-dependent stellar lifetime (\citealt{Portinari98};
\citealt{Marigo01}) and the assumed IMF.

A top heavy IMF in bursts increases the number of supernovae that
explode per unit of mass turned into stars, thereby generating
stronger feedback. This, in turn, inhibits the early collapse of cold
gas clouds in small subgalactic halos, helping to maintain an abundant
reservoir of hot halo gas. Following the last major merger and
accompanying starburst, radiative cooling of hot gas from the 
reservoir flows inwards and settles into a centrifugally supported
disc which becomes unstable to star formation. This is the key to the formation of
a realistic disc galaxy in the simulations of \cite{Okamoto05}. 

\subsection{Identifying central and satellite galaxies}
\label{Identify} 

We begin the process of finding central and satellite galaxies by
identifying ``friends-of-friends'' (FOF) groups in the dark matter, 
linking together particles whose separation is less than 0.2 times the
mean interparticle separation, corresponding roughly to particles
within the virialised halo (\citealt{Davis85}). We then identify bound
substructures (``subhalos'') in the simulation with the algorithm
\texttt{SUBFIND} (\citealt{Springel01}). Using particle
positions and velocities, \texttt{SUBFIND} calculates the binding
energy of each FOF group (stripping unbound particles of group
membership) and then identifies self-bound substructures inside the parent
halo. We consider only halos with ten or more dark matter particles,
corresponding to a subhalo mass resolution of $M_{\rm res, sub}=
1.7\times10^{8} h^{-1} M_{\odot}$. For each subhalo, we calculate the
centre of mass, as well the size, $r_{\rm sub}$, defined as the rms
distance of its particles from the centre.

Substructures within the parent halo are thus identified as locally
overdense, self-bound, regions in the dark matter density field that
fall within the high resolution region of the simulation. In order to
identify clumps of star particles as individual galaxies, we associate
each star particle with a unique dark matter substructure. For each
gas and star particle, we find the substructure whose centre of mass
is closest and assign the SPH particle to that substructure if it is
within $r_{\rm sub}$. Since star clumps tend to be dense and centrally
concentrated, our results are robust to the choice of subhalo radius
(for example, increasing $r_{\rm sub}$ by 50\% has a negligible effect on
our results). We also impose a ten star particle lower limit,
corresponding to $M_{\rm res, gal} \sim 1\times10^{7}
h^{-1}$M$_{\odot}$, on the stellar mass of any galaxy. With this
resolution limit, our galaxy samples become incomplete at magnitudes
fainter than $M_{V} \sim -12$.

We identify ``central'' galaxies of luminosity similar to the Milky
Way with galaxies brighter then -20.4 in the V band. We calculate
their virial radius by growing concentric spheres around the galaxy
and noting where the mean internal density first falls below 200$\rho_{\rm
crit}$, where $\rho_{\rm crit}$ is the critical density.  Galaxies
that fall within this radius are considered satellite galaxies. In
order to obtain isolated systems similar to the Milky Way, we ensure
that no two central galaxy candidates are within each other's virial
radius.
\begin{figure}
\includegraphics[width=20pc]{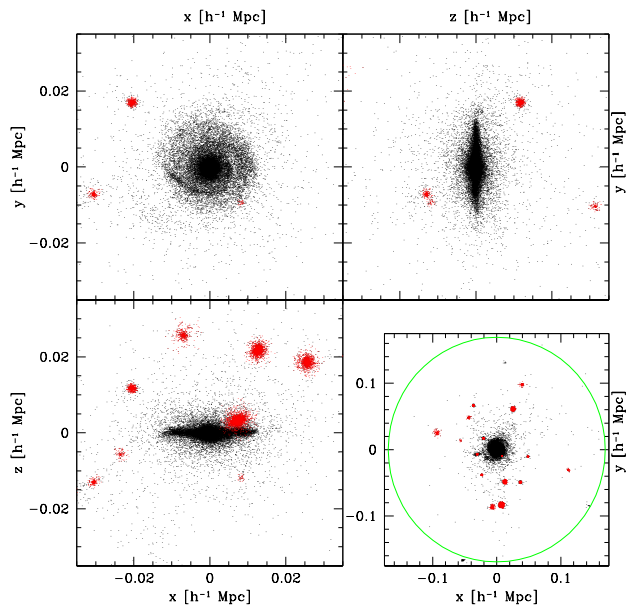}
\caption{A central galaxy and its satellite system. The black dots
denote all the star particles associated with the galaxy, while the
red dots denote the star particles associated with satellites located
within the virial radius of the central galaxy. The top two and the
bottom left panels show a zoomed-in projection of the distribution of
stars near the central galaxy, in the three principal planes, i.~e. in
the planes in which the galaxy is face-on and edge-on. The {\it bottom
right} panel shows a face-on projection out to $250$~kpc. The virial
radius, defined by $\bar
\rho(<r_{\rm vir})= 200 \rho_{\rm crit}$, is marked by the green circle.}
\label{fig1}
\end{figure}

Fig.~\ref{fig1} shows an illustrative example of how our algorithm
selects central and satellite galaxies. We plot all the star particles
within the virial radius projected onto the principal planes of the
central galaxy, i.e. the planes on which the galaxy is face-on and
edge-on. In red, we plot the star particles that are associated with
satellite galaxies. There are different numbers of satellites in each
projection because we are plotting all the star particles in
projection through the virial radius, which is larger than the plotted
box. In the bottom right panel, we show the face-on projection of the
system out to the virial radius. The central galaxy has a very
extended stellar halo (black points in the figure) made up of stars
that lie within the $r_{\rm sub}$ of the central galaxy' subhalo and
not within the $r_{\rm sub}$ of any satellite subhalo. This extended
halo makes up only a small fraction of the stellar mass of the central
galaxy. Below, when we calculate the shape of the central galaxy we
remove the extended halo by considering only the innermost 98\% of the
stars.
\begin{table*}
\label{properties}
\begin{center}
 \caption{Properties of the 9 galaxy haloes (gh). We show the virial
 radius ($r_{\rm vir}$), the dark halo mass internal to this radius
 ($M_{\rm dm}$), the stellar mass
 of the galaxy ($M_{\star}$), the gas mass of the galaxy ($M_{\rm
 gas}$), the V band magnitude ($M_{V}$), the
 number of satellites orbiting within the virial radius ($N_{\rm
 sat}$), and the bulge ($M_{\rm bulge}$) and disc ($M_{\rm disc}$) mass
 of the galaxy.}
 \end{center}
 \begin{center}
   \begin{tabular}{l l l l l l l l l l l}
         & gh1 & gh2 & gh3 & gh4 & gh5 & gh6 & gh7 & gh8 & gh9\\
   \hline
   \hline
        $r_{\rm vir}$ ($h^{-1}$ Mpc) & 0.171 & 0.170 & 0.169 & 0.146 & 0.106
	     & 0.133 & 0.120 & 0.106 & 0.151 \\
	$M_{\rm dm} (h^{-1} 10^{12}M_{\odot})$ & 1.170 & 1.154 & 1.127 & 0.731
        & 0.281 & 0.458 & 0.397 & 0.284 & 0.795 \\
	$M_{\star} (h^{-1} 10^{10}M_{\odot})$ & 4.64 & 6.73 & 4.54 & 3.63
        & 1.62 & 2.28 & 1.69 & 1.21 & 3.88\\
	$M_{\rm gas} (h^{-1} 10^{10}M_{\odot})$ & 2.19 & 3.51 & 3.78 & 1.86 &
        0.54 & 1.86 & 1.22 & 0.89 & 2.40 \\
	$M_{V}$ & -21.3 & -22.1 & -21.2 & -20.9 & -20.4 & -20.5 & -20.4 &
        -20.4 & -21.6\\
	$N_{\rm sats}$ & 20 & 13 & 16 & 4 & 5 & 5 & 8 & 5 & 2\\ 
	$M_{\rm bulge} (h^{-1} 10^{10}M_{\odot})$ & 2.64 & 1.89 & 1.76  &
        1.78 & 0.71 & 1.30 & 1.04 & 1.01 & 2.01\\
	$M_{\rm disc} (h^{-1} 10^{10}M_{\odot})$ & 2.00 & 4.84 & 2.79 &
        1.85 & 0.92 & 0.99 & 0.66 & 0.20 & 1.87\\

    \hline
    \hline
 \end{tabular}
 \end{center} 
\end{table*}

\section{The luminosity function of satellite galaxies}
\label{luminosity} 

Our simulations produced a sample of 9 central galaxies with
$M_{V}<-20.4$, 8 from simulation~SR and one from simulation~SD. 
The properties of these 9 systems are displayed in Table~1.

\begin{figure}
\includegraphics[width=20pc]{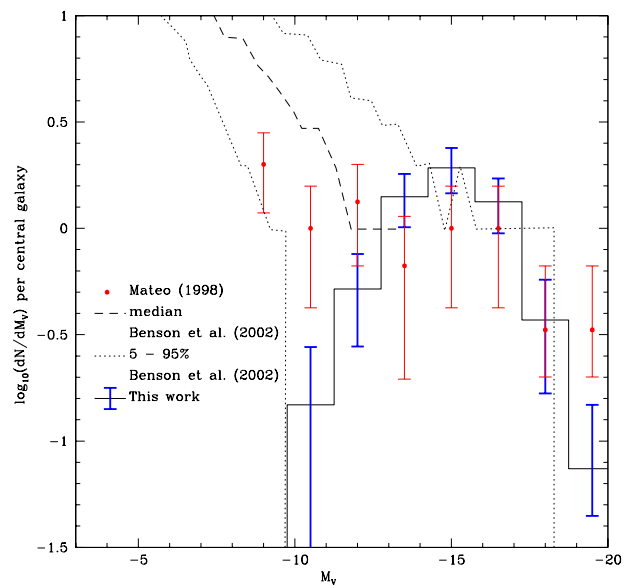}
\caption{The differential V-band luminosity function of satellites per
central galaxy in our simulations. The blue error bars are $1\sigma$
Poisson errors per magnitude bin. (For bins with more than one count
the errors are simply $\sqrt{N}$, but for bins with only one count, we
take $+\sqrt{N}$ as the upper errorbar and determine the lower
errorbar by finding the mean of a Poisson distribution whose integral
from 1 to infinity corresponds to 16\% of the distribution.) The Local
Group satellite luminosity function, obtained from the compilation of
Mateo (1998), supplemented with data from Irwin, is shown as
red circles. The dashed line shows the median luminosity function of
70 realisations of Milky Way type halos calculated using a
semi-analytic model by Benson et al (2002b) with the dotted lines
indicating the $5$ -- $95$ percentile width of their distribution.}
\label{satgals_lf}
\end{figure}

The V-band luminosity function of the satellites in this sample is
shown in Fig.~\ref{satgals_lf}. The luminosity function is nearly flat
in the range $-13.5>M_{V}>-16.5$ and drops off sharply at fainter and
brighter magnitudes. The nominal resolution limit of our simulations
corresponds to a satellite absolute magnitude $M_V\sim -12$ and it is
important to check whether the decline in the luminosity function in
the range $-12>M_{V}>-13.5$ is the result of feedback processes, such
as photoionization or supernova feedback, which affect the number of
faint galaxies in the simulations, or whether it is due to limited
resolution.

To investigate the likely effects of resolution in our estimate of the
satellite luminosity function we have used the semi-analytic model
\texttt{GALFORM} described by \cite{Cole00} and \cite{Benson02a}. The
semi-analytic model includes all the standard physical effects present
in the simulation: gas cooling according to a metallicity-dependent
cooling function, star formation, feedback due to supernovae
explosions, etc. Photoionization is included in an approximate way by
assuming that gas cannot cool in halos with circular velocity $v_{\rm
circ} < 60$~km~s$^{-1}$ after the assumed epoch of recombination,
$z<6$. \cite{Baugh05} have shown that this simple approximation gives
an excellent match to a detailed calculation based on the concept of a
filtering mass (\citealt{Gnedin00}, \citealt{Benson02b}).

We proceed as follows.  In the semi-analytic model, it is possible to specify
the mass of the smallest halo to be considered and this allows us to model the
effects of resolution in the SPH simulation (\citealt{Helly03}). We recall that
in the simulation itself we only considered halos with at least ten particles,
corresponding to $M_{\rm res, sub}=1.7 \times 10^8 h^{-1}$~M$_\odot$. We ran a series
of semi-analytic models with resolution varying by 4 orders of magnitude, from
$\sim10^{-2}M_{\rm res, sub}$ to $10^{2}M_{\rm res, sub}$. In all cases, we
found that the shape of the luminosity function brightwards of $M_{V}=-12$ was
essentially unaffected. If the resolution is degraded further, then the faint
end of the luminosity function becomes truncated at increasingly bright
magnitudes. We conclude from this test that our satellite sample is likely to be
complete for magnitudes brighter $M_{V}=-12$ and that our estimate of the
satellite luminosity function in this regime is unlikely to be affected by
resolution.

In Fig.~\ref{satgals_lf}, we compare our estimated luminosity function
with data for satellites in the Local Group obtained from the sample
compiled by \cite{Mateo98}, supplemented with data from
Irwin\footnote{http://www.ast.cam.ac.uk/$\sim$mike/local\_members.html}.
Unlike \cite{Mateo98}, we include the SMC, the LMC and M33 in our
sample.  Brighter than our estimated resolution limit of $M_{V}<-12$,
we find that the luminosity function in our simulations is in
excellent agreement with the Local Group data over a range of 7 magnitudes, down
to $M_{V}=-19$. Of course, in the Local Group, the statistics at the
bright end are rather poor: the last two data points (centred on $-18$
and $-19.75$) contain only one galaxy each (the LMC and M33,
respectively). However, within the errors, the simulations are
consistent with the data.

We also compare our results in Fig.~\ref{satgals_lf} with those in the
semi-analytic model of \cite{Benson02b}. The simulations and the
semi-analytic model are broadly in agreement over most of the
luminosity range, from the resolution limit of the simulations to the
brightest two bins. However, there is a significant difference at the
bright end: while we find a satellite as bright as the LMC one third
of the time, Benson et al. only find one such satellite 5\% of the
time. In the simulations, satellites as bright as M33 are produced in
about 5\% of systems, while the frequency in the Benson et al. model
is less than 1\%. The reasons for the disagreement between the
semi-analytic model and the simulations (and Local Group data) are
a reflection of the different treatment of various physical
processes in the methods. However, the fact that our simulations match
the observed satellite luminosity function over a large range of
magnitudes indicates that the relatively small observed number of
satellites in the Local Group is not, in principle, difficult to
explain within the CDM model. The much publicized ``satellite
problem'' exists, as other authors have remarked
(e.g. \citealt{Bullock00,Benson02b}), only when the astrophysical
processes involved in the formation of visible galaxies are neglected.

\begin{figure}
\includegraphics[width=20pc]{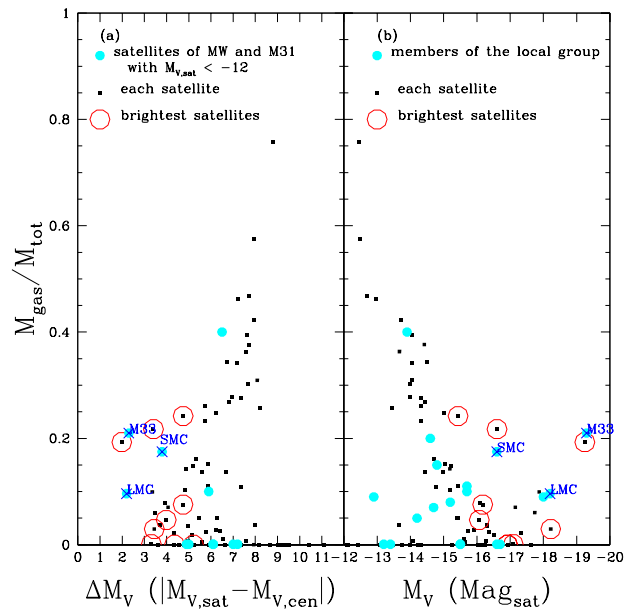} 
\caption{Gas fraction
in satellites as a function of V-band luminosity.  The satellites in
the simulation are shown as squares and the brightest satellite in
each system is surrounded by a red circle. The Local Group satellites
are shown as filled cyan circles and the LMC, SMC and M33 are indicated by
crosses and labelled. \textit{Panel (a)}. Gas fraction as a function
of the magnitude difference between the satellite and its central
galaxy. For both simulations and data, only satellites with
$M_{V}<-12$ that lie within the virial radius of the central galaxy
(the Milky way or M31 in the case of the real data) are plotted; in
addition, for the real data a reliable gas fraction measurement or
upper limit is required.  
\textit{Panel (b)}. Gas fraction as a function of absolute
V-band magnitude. Here, all Local Group members with reliable gas
fraction measurements or upper limits are plotted.  Local Group data
are from Mateo (1998); SMC data are from Stanimorovi, Staveley-Smith \&
Jones (2003); LMC data are from Staveley-Smith et al. (2003); and M33
data are from McGaugh (2005).} 
\label{gas_content} 
\end{figure}

In order to investigate how realistic the satellites in our
simulations are, particularly the bright ones, we compare, in
Fig.~\ref{gas_content}, their gas fractions with those of real
satellites.  
Panel (a) shows the gas fractions as a function of the magnitude
difference between the satellite and the parent galaxy in the
V-band. Only satellites with $M_{V}<-12$ in the simulations, the Milky
Way and M31 are shown and, for the latter, we require also a reliable
measurement of, or upper limit to the gas fraction. The simulations
show a trend of increasing gas fraction with decreasing luminosity
which, as far as the scant data for the real satellites permit, seems
consistent with the measurements. In particular, the brightest
satellites in the simulations bracket the values measured for the LMC,
the SMC, and M33 with 3 out of 9 simulated satellites having a larger
gas fraction than the LMC and SMC. In panel (b) we extend the
comparison by including not only satellites within the virial radius,
but all the satellites in the Local Group that have reliable gas
fraction measurements or upper limits. In this case, we plot the gas
fraction against the V-band magnitude of the satellite. The trend of
increasing gas fraction with decreasing luminosity is now clearer in
the data and the locus they define agrees well with the locus traced
by the simulated satellite galaxies (although note that not all the
points plotted are strictly speaking satellites, according to our
definition, which requires satellites to lie within the virial
radius.) We conclude that the satellites in the simulations not only
have a luminosity function similar to that observed, but also have
realistic gas fractions. 

\section{The spatial distribution of satellite galaxies}
\label{alignments} 

\begin{figure}
\includegraphics[width=20pc]{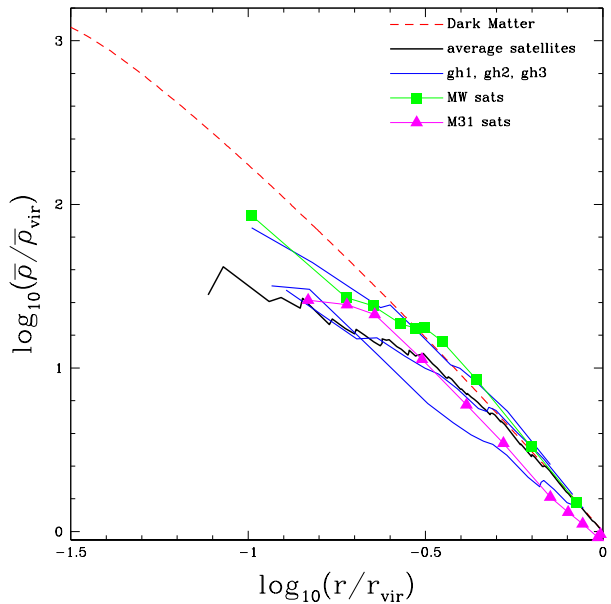} 
\caption{Mean interior radial profiles for dark matter and satellites.
The halo dark matter density profile, averaged over the 9 systems with
central galaxies, is shown by a red dashed line. The number density
of satellites, also averaged over the 9 systems, is shown by a solid
black line. The three blue lines show the number density profiles of
satellites for the three systems with 11 or more satellites (gh1, gh2
and gh3) and may be compared with the number density profiles for the
11 brightest satellites in the Milky Way (green squares and lines)
and in M31 (magenta triangles and lines). In all these cases, the
satellites were ordered according to distance from the centre and the
density was evaluated at the mid point of the logarithm of the distance
between adjacent satellites.  The radial coordinate is in units of the
virial radius, $r_{\rm vir}$, and all profiles have been normalised to
the mean interior density at the virial radius.}
\label{dplot} 
\end{figure}

In Fig.~\ref{dplot}, we show the mean interior radial mass density
profile of the dark matter (red dashed line) and the mean interior
radial number density profile of the satellites (thick black line)
averaged over the 9 central galaxies in our simulations.
Also plotted are the individual number density profiles of the
satellites in the three systems with 11 or more satellites (gh1,
gh2 and gh3; blue lines), as well as the corresponding profiles for
the 11 brightest satellites in the Milky Way (magenta line) and M31
(green line).  All profiles have been normalized to their respective
values at the virial radius. The dark matter density profile, shown as
the dashed line, closely follows the NFW form (\citealt{NFW1996};
\citealt{NFW1997}).  As found in previous N-body studies
(\citealt{Stoehr02}; \citealt{Gao04}; \citealt{Libeskind06}), the
average satellite number density profile is flatter than the dark
matter mass density profile. The three simulated systems with 11 or
more satellites display a range of profiles, but they all resemble the
observed profiles of real satellites, and one of them (gh2), is
remarkably close to the Milky Way data over the entire radial range,
0.1-1$r_{\rm vir}$.

To characterize our simulations further, we investigate the flattening
of the distributions of dark matter, satellites and stars in the
central galaxy. We define the tensor of second moments,
\begin{equation} 
\label{Inerteq} 
I_{jk}=\sum_{\mu}x_{j}^{\mu}x_{k}^{\mu}, 
\end{equation}
where $x_{j}^{\mu}$ is the $j$ coordinate of the $\mu$th particle in a
reference frame centred on the centre of mass of the central
galaxy. To determine the flattening of the central galaxy, we consider
only the innermost 98\% of the stars and exclude the outermost 2\%
that make up the diffuse stellar halo.  (Our results are insensitive
to the exact fraction of excluded stars, provided this is small.) To
determine the principal axes of each distribution, we diagonalise the
tensor $I_{jk}$. Its eigenvalues, $a^{2}, b^{2}, c^{2}$, give the mean
square deviation of the $x$, $y$ and $z$ coordinates along the
principal axes. We define $a > b > c$ as the major, intermediate and
minor axes of the system, and use the notation \{$a_{\rm dm}, b_{\rm
dm}, c_{\rm dm} $,\}, \{$a_{\rm sat}, b_{\rm sat}, c_{\rm sat}$\},
\{$a_{\rm gal}, b_{\rm gal}, c_{\rm gal}$\} to refer to the
distributions of dark matter, satellites and stars in the
central galaxy respectively.

It is important to keep in mind that estimates of $a$, $b$ and $c$
using small numbers of objects are generally biased towards higher
anisotropy. For example, if a system has only three members
(e.g. satellite galaxies), our method would always return $c=0$ since 
three objects will always lie on a plane. Tests of the statistical
robustness of our results are performed below.

\begin{figure}
\includegraphics[width=20pc]{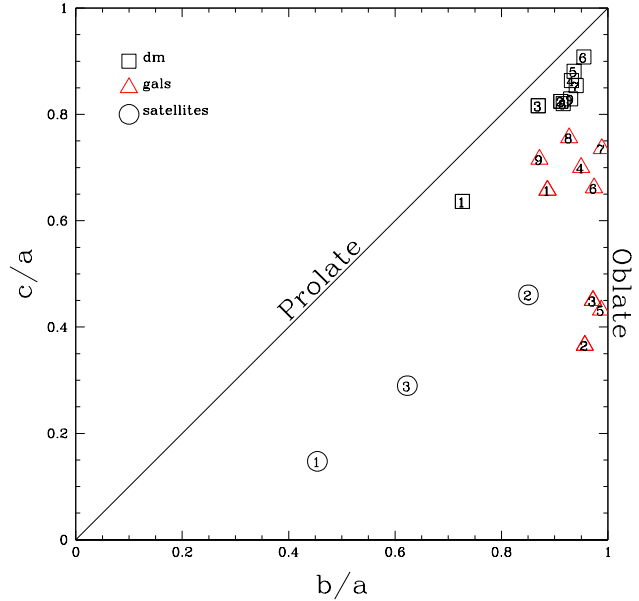}

\caption{Minor-to-major ($c/a$) versus intermediate-to-major ($b/a$)
axial ratios for all the galactic systems in our simulations. Since
$a>b>c$, the upper left triangular half of this plot cannot contain
any points. Prolate objects lie along the diagonal while oblate
objects lie along the right vertical axis. The numbers inside each
symbol identify the galaxy. Axial ratios for stellar discs and for
dark matter halos are shown for all nine systems by triangles and
black squares respectively. For the three systems with 11 or more 
members, the axial ratios of the structure defined by the 11 most
massive satellites are shown by the circles.}
\label{cavba}
\end{figure}

The Milky Way has 11 satellite galaxies within its virial radius with
reliably measured positions and magnitudes. As shown in Table~1, of
our sample of nine simulated central galaxies, three have 11 or more
satellites and can be compared with the Milky Way. The axial ratios of
each central galaxy in our simulations, its dark matter halo and, for
systems with 11 or more members within the virial radius, its
satellite distribution are shown in Fig.~\ref{cavba}.

It is immediately apparent that, in the three sufficiently rich
systems, the distribution of the 11 most massive satellites is
considerably flatter than their respective dark matter halo,
particularly in systems~1 and~3. This is consistent
with previous studies (\citealt{Libeskind05}, \citealt{Kang05},
\citealt{Zentner05}) which also found large flattenings in simulated
satellite systems, although in those studies the satellites were
identified in dark matter simulations using a semi-analytic galaxy
formation model rather than the full hydrodynamic calculations that we
are analyzing here.  Libeskind et al ascribed this anisotropy to the
preferential infall of substructures along the spines of filaments as
they collapse to make a galaxy.

In order to assess the statistical robustness of our results, 
we performed three tests whose results are displayed in Table~\ref{prob}. 
\begin{table}
\begin{center}
 \caption{The probability of randomly drawing values of
 the minor-to-major ($c/a$) and intermediate-to-major ($b/a$) axis
 ratio which are more extreme than those 
 measured for the satellite systems in the simulations according to the
 three test described in the text. The three halos with 11
 or more satellites within the virial radius are labelled gh1, gh2
 and gh3 respectively.}
\end{center}
\begin{center}
\begin{tabular}{l l l l l l}
        & gh1 & gh2 & gh3\\
   \hline
   \hline
     Isotropic NFW sphere   & 0.1\%  & 47.6\%  & 5.4\% \\
     Squashed NFW profile   & 0.4\%  & 51.5\%  & 6.4\%  \\
     Simulated dark halo & 0.6\%  & 66.2\%  & 10.5\% \\
    \hline
    \hline

 \end{tabular}
 \end{center} 
\label{prob}
\end{table} 
For the first test, we constructed a spherically symmetric NFW halo with
$10^{6}$ particles. We then selected 11 particles at random from the
halo 1000 times, and calculated the cumulative distributions of
minor-to-major and intermediate-to-major axis ratios. The probability
of drawing more extreme values that those measured for the satellite
distributions in the simulations is given in the first row of 
Table~\ref{prob}, for each of the three galaxies with 11 or more
satellites. For the second test, we construct flattened NFW
halos by squashing a sphere according to the values of $a_{\rm dm}$,
$b_{\rm dm}$, and $c_{\rm dm}$ for each of the three simulated halos.
We then performed the same test as before; the results are displayed in
the second row in Table.~\ref{prob}. Finally, we selected 11 dark matter
particles directly from each of the three halos (again 1000 times) and
performed the same test with results given in the third row of the
table.

\begin{figure*}
\hbox{
\includegraphics[width=22pc]{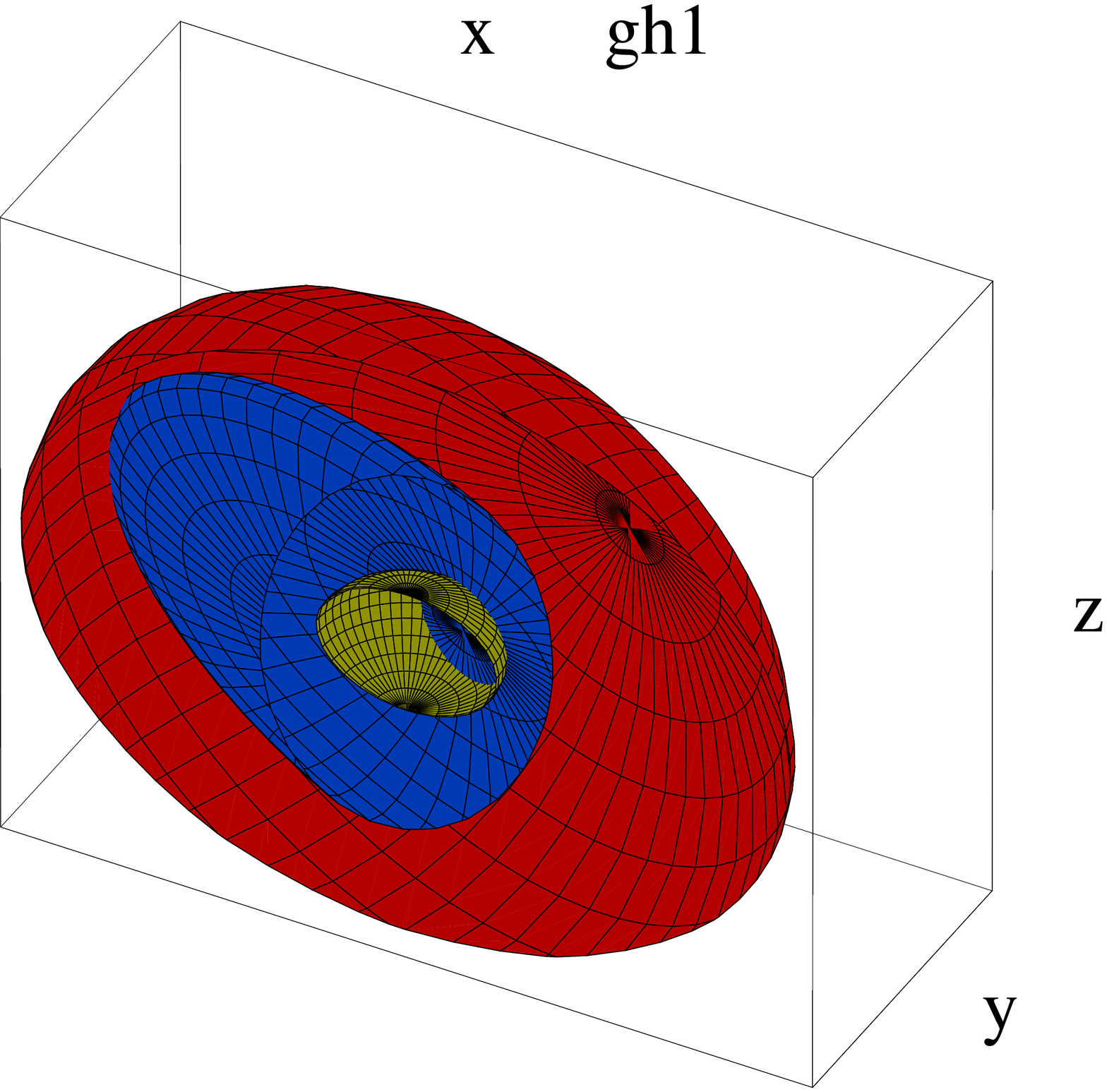}
\includegraphics[width=22pc]{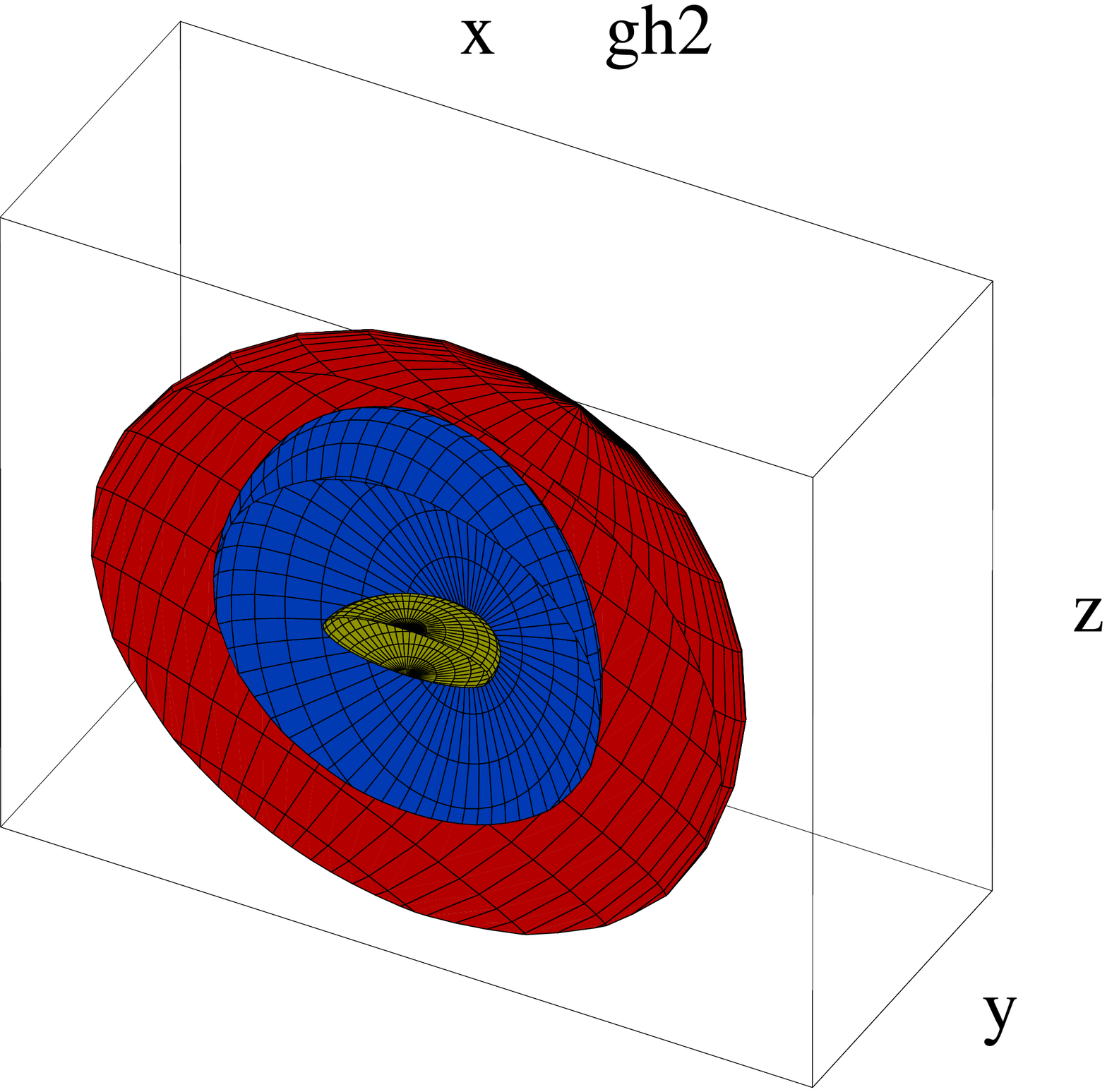}}
\includegraphics[width=22pc]{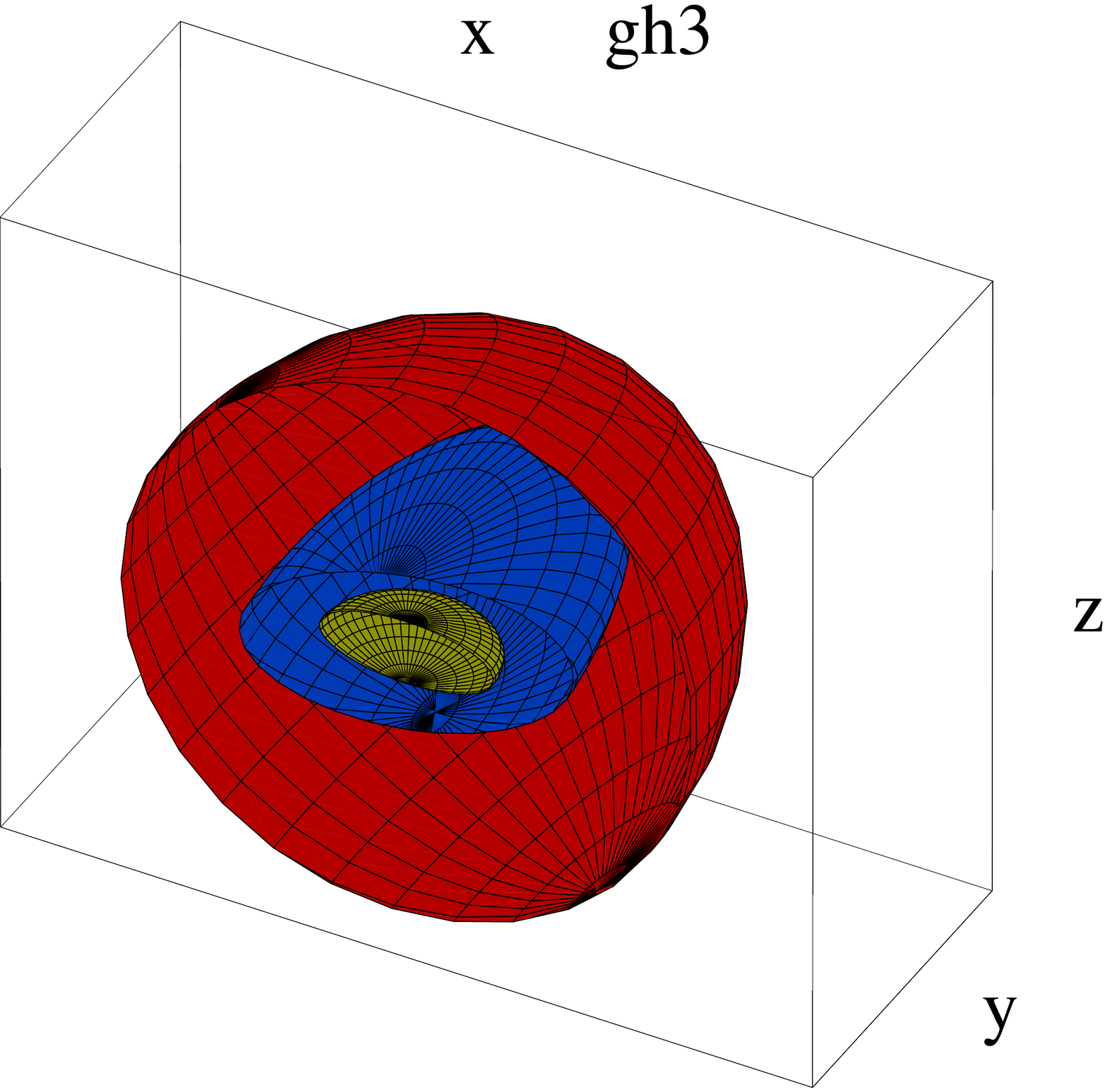}
\caption{The three simulated galaxies with 11 or more
satellites. Clockwise, from the top-left, the plots show galaxies
gh1, gh2 and gh3 respectively.  The central galaxy, the satellite
system and the dark matter halo are represented by the yellow, blue
and red ellipsoids respectively. While each ellipsoid has been
flattened and oriented according to the measured parameters, their
relative sizes have been arbitrarily chosen for clarity. Note that the
long, medium and short axes of the central galaxy are always aligned
with the $x$, $y$, and $z$ axis.}
\label{3drends}
\end{figure*}  

With the exception of galaxy gh2, whose satellite system is the least
flattened of the three (see Fig.\ref{cavba}), the numbers in
Table~\ref{prob} rule out with high confidence the null-hypothesis
that the satellite systems in the simulations were randomly drawn from
either a spherical or a squashed NFW halo or from the dark matter
distribution of the halo itself.  This result agrees well with those
obtained by \cite{Libeskind05} and \cite{Zentner05}.

The flattening and relative orientations of the disc, halo and
satellite system of each of our three well populated simulations are
shown schematically in cross section in Fig.~\ref{3drends}.  While the
relative flattening and orientation of each component has been
rendered according to the measured values, for clarity, the relative
sizes have been chosen arbitrarily. The central galaxy is represented
by a yellow ellipsoid embedded within two larger ellipsoids, a blue
one representing the satellite system and a red one representing the
dark halo. The poles (i.e. the direction of the $\hat{c}$ axis) are
visible as the point at which the lines of ``longitude''
converge. The various systems display a variety of orientations
which we quantify below.

The central galaxy orientations displayed in Fig.~\ref{3drends} were
calculated from the innermost 98\% of stars. To refine our estimate of
orientation and for other purposes, we have performed a dynamical
bulge-disc decomposition for each of the nine central galaxies in our
simulations, applying the method proposed by \cite{Abadi03} and used
also by \cite{Okamoto05}. The net angular momentum, relative to the
centre of mass of the galaxy, of the material in the inner $10
h^{-1}$~kpc is used to define a ``z'' axis. The angular momentum,
$J_{\rm z}$, about this axis is computed for each star particle and
compared to the angular momentum of a circular orbit with the same
energy, $J_{\rm c}(E)$.  All stars with $J_{\rm z}/J_{\rm c}(E) \geq
0.75 $ are assigned to a disc component whose orientation is taken to
be the direction of the nett angular momentum of the disc
stars. (In some cases, this method may incorrectly assign some bulge stars
to the disk, but this is not important for our analysis.)
Generally, the direction defined by the disc angular momentum coincides with
that of the short axis of the overall stellar distribution, except in
two case (galaxies gh8 and gh9) which are bulge-dominated galaxies
with very small discs whose orientation is ill-defined. In what
follows, we take the orientation of the galaxy to be that of the disc
except for gh8 and gh9 for which we take the direction of their minor
axis.

\begin{figure} \includegraphics[width=20pc]{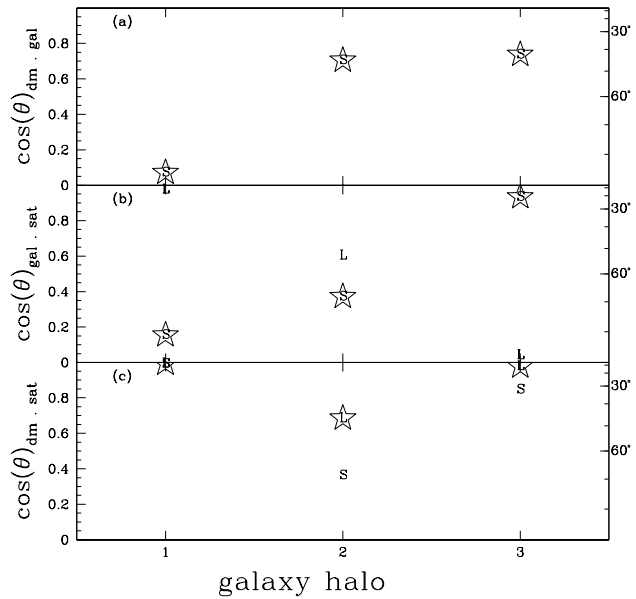}
\caption{Alignments between the long (L) and short (S) axes of the
different components of the three galaxies with 11 or more
satellites. \textit{Panel (a):} The cosine of the angles 
between the axes of the dark halo and of the 
galaxy. \textit{Panel (b):} The cosine of the angles between the
axes of the galaxy and of the satellite distribution.
\textit{Panel (c):} The cosine of the angles between the axes of
the dark halo and of the satellite distribution. The right-hand
coordinates are labelled with the value of the angle. The most
interesting relations are highlighted with stars.}
\label{align.shape}
\end{figure}

The angles between the different components of each of the three
galaxies with 11 or more satellites are plotted in
Fig.~\ref{align.shape}.  The top panel shows that in two of the
galaxies, the disc is inclined about $\sim 45^\circ$ relative to the
minor axis of the dark halo. Surprisingly, in the third galaxy, the
disc is orthogonal to the minor axis of the halo.  The middle panel
shows that in two cases (including the one with the orthogonal disc),
the satellite systems are, within $20^\circ$, perpendicular to the
galactic disc.  This is also surprising but it is exactly the
alignment seen in the Milky Way galaxy, where the satellites lie
approximately along a great circle on the sky whose pole is in the
galactic plane (e.g. \citealt{Kroupa05}). In the third system (gh3),
the satellite system is almost perfectly aligned with the galactic
disc. Finally, the bottom panel shows that in two of the three
cases, the long axis of the satellite system is well aligned with
the long axis of the dark halo. This is consistent with the
conclusions reached by \cite{Libeskind05} from their (discless) dark
matter simulations.  The lack of alignment in system gh2 is probably
due to the fact that this has the least aspherical dark halo whose
long axis is therefore poorly determined.  

\section{Alignment of the angular momenta}
\label{angmom} 

In this section, we extend our study of alignments to include the
possible correlations between the angular momenta of the dark matter halo,
$\Jdm$, the satellite galaxy population, $\Jsat$, and the stars of the
central galaxy, $\Jdisc$.  Unless otherwise noted, we consider a
number-weighted mean angular momentum, $\underline{\rm J}= \langle
\underline{r} \times \underline{v} \rangle$. For the dark matter halo
and the central galaxy whose particles all have roughly the
same mass, this ``specific'' angular momentum is approximately
proportional to the
standard, mass-weighted angular momentum ($\langle \underline{r}
\times \underline{p} \rangle$). For a satellite system, however,
this definition gives equal weight to each satellite, preventing the
statistics from becoming dominated by one or two very large
satellites.  

\begin{figure}
\includegraphics[width=20pc]{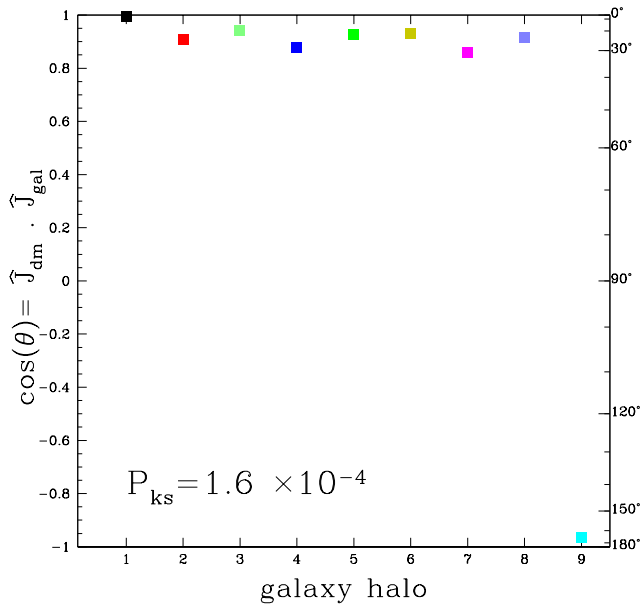}
\caption{Alignment of the angular momentum vectors of the galaxy
($\Jdisc$) and the dark halo ($\Jdm$) for material within 
the virial radius for our nine simulated galaxy systems. The
right-hand axis gives the value of the angle. Each galactic
system has been assigned a colour that will be retained throughout the
remainder of this paper. }
\label{align.all}
\end{figure}

The angle between $\Jdm$ and $\Jdisc$ evaluated for all the material
within the virial radius, is plotted in Fig.~\ref{align.all}. Since
the number of satellites is immaterial in this case, we expand our
sample to include the six other galactic systems (identified according
to the criteria outlined in Sec.~\ref{Identify}) that were excluded
from the satellite analysis in the preceeding section. The angular
momenta of the galaxy and the dark matter within the virial radius
are aligned to within $\sim 30^\circ$, i.e. the galaxy spins in
essentially the same direction as the dark matter halo.  The single
exception is system gh9 in which the galaxy and the dark matter are
counter-rotating.  A KS test shows that the probability of obtaining
the distribution of cos$\theta$ shown in the figure from nine objects
drawn at random from a larger sample of randomly oriented galaxies is
only $\sim 0.016 \%$. We conclude that there is a significant
alignment between the angular momenta of the galaxy and that of the
dark halo it inhabits. This is consistent with the acquisition of
angular momentum prior to the collapse of the system, when dark matter
and gas were well mixed, as expected in the tidal torque theory
(e.g. \citealt{White84}) and as usually assumed in semi-analytic
models of galaxy formation (e.g. \citealt{Cole00}).

\begin{figure}
\includegraphics[width=20pc]{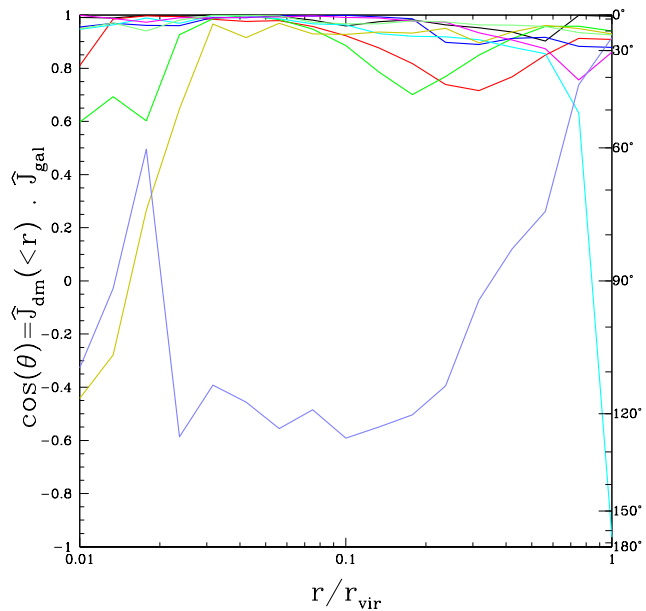}
\caption{The angle between the angular momentum of the disc
($\underline{\rm J}_{\rm disc}$) and the dark matter halo ($\Jdm(<r)$)
as a function of radius. Each colour represents a single system
according to the same code used in Fig~\ref{align.all}.}
\label{alignfr}
\end{figure}

The radial dependence of the galaxy-halo spin alignment is shown in
Fig.~\ref{alignfr} where we plot the cosine of the angle between
$\Jdisc$ and $\Jdm (< r)$ for our nine systems.  In 7 out of 9 cases,
the alignment persists out to the virial radius of the halo, never
straying much beyond $\sim 30^\circ$.  The two exceptions are systems gh9
and gh8. In the first of these, a tight alignment persists for
$r<0.7r_{\rm vir}$, but it rapidly disappears beyond this radius. A
visual inspection of this system shows that the reason for the rapid
change in spin direction is simply the presence of a large
counter-rotating fragment of dark matter in the outer parts of the
halo which has recently been accreted, causing the halo angular
momentum to flip in the outer parts. The other anomalous system owes
its strange behaviour to the fact that the galaxy is almost a pure
spheroid with very little angular momentum and for which the 
direction of $\Jdisc$ is poorly defined.

Previous studies (e.g. \citealt{Warren92}, 
\citealt{BailinSteinmetz05}) have found that the halo angular
momentum, $\Jdm$, tends to be aligned with the halo minor axis,
$\cdm$.  The simulations of \cite{Warren92} clearly show such an
alignment in the inner parts of the dark halo. Bailin \& Steinmetz
(2005; see their Fig.~16) demonstrate that the alignment becomes
weaker with radius but is still significant ($\sim 30^{\circ}$) at the
virial radius. By contrast, \cite{Bailin05} report an absence of
alignment between $\cdm$ and $\Jdisc$ at the virial radius in seven
hydrodynamic simulations. 
We now consider the angle subtended by $\Jdm$ and
$\cdm$  in our own simulations.
\begin{figure}
\includegraphics[width=20pc]{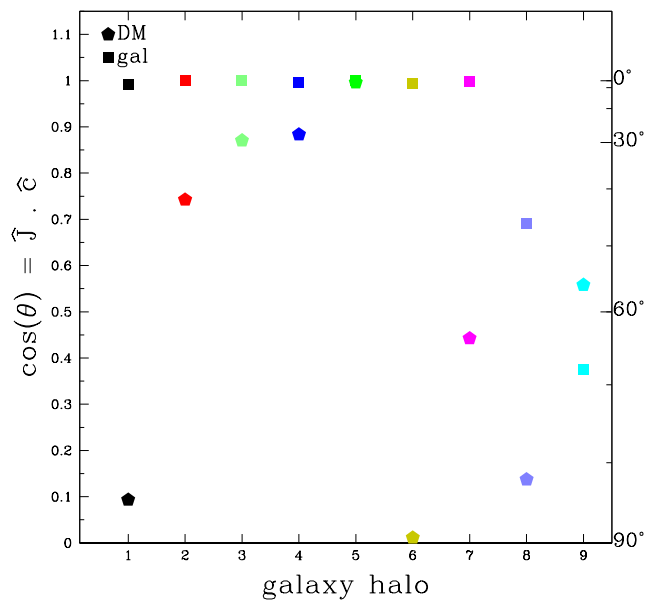} \caption{The angle
between the angular momentum and the short axis of the dark matter
halo ($\Jdm . \cdm$, pentagons) and of the galaxies ($\Jdisc . \cdisc$,
squares) for the nine galactic systems. The angular momentum vector
in most of the galaxies points in the direction of the shortest
galactic axis. For the halos, however, no such correlation exists. }  
\label{Jdmcdm}
\end{figure} 

\begin{figure}
\includegraphics[width=20pc]{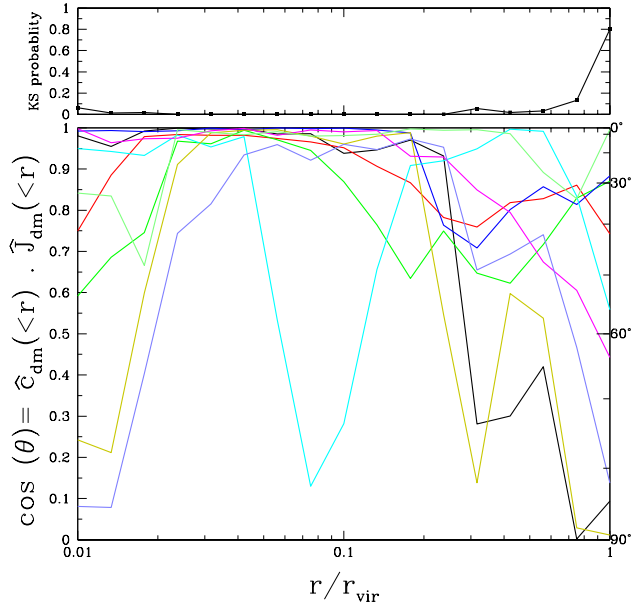} 
\caption{The angle between $\Jdm$ and $\hat{\rm c}_{\rm dm}$ for
material within a given radius. Each of the 9 galactic halos is
plotted using a different colour, according to the same code used in
Fig.~8.}
\label{Jdmcdmfr} 
\end{figure}

In Fig.~\ref{Jdmcdm} we plot the angle between $\Jdm$ and $\cdm$ for
all the material within the virial radius and, in Fig.~\ref{Jdmcdmfr},
the run of this angle with radius. We also plot the KS probability
that the distribution of angles in the nine systems at each radius is
consistent with a uniform distribution.
Our results are qualitatively in good agreement with both
\cite{Warren92} and \cite{BailinSteinmetz05}.  The radial dependence
of $\Jdm . \cdm$, displayed in Fig.~\ref{Jdmcdmfr}, shows that, apart
from one anomalous system, gh9, there is a good
alignment between these two vectors in the inner parts of the halo,
within a few tenths of $r_{\rm vir}$.  In the outer parts the 
correlation becomes much weaker, which is perhaps not surprising for
systems whose shape is supported by an anisotropic velocity dispersion
tensor rather than by rotation (\citealt{Frenk85,Warren92}). While
\cite{Warren92} and \cite{BailinSteinmetz05} find a weak correlation
at the virial radius our small sample of 9 halos is consistent with no
correlation.  Thus, the KS probability that the cosine of the angle for
these 9 halos is consistent with a uniform distribution jumps from
$\sim 15 \%$ at $>0.7r_{\rm vir}$ to $\sim 80\%$ at $r_{\rm vir}$.

Turning to the galaxies, Fig.~\ref{Jdmcdm} shows that in all but two
of them, the angular momentum vector points in the direction of the
shortest galactic axis, as expected for systems flattened by
rotation. Of the two discrepant galaxies, gh8 is a nearly perfect
spheroid with very little rotation while gh9 is a rapidly rotating
spheroid, whose angular momentum is dominated by the bulge's bulk
rotation.

\begin{figure}
\includegraphics[width=20pc]{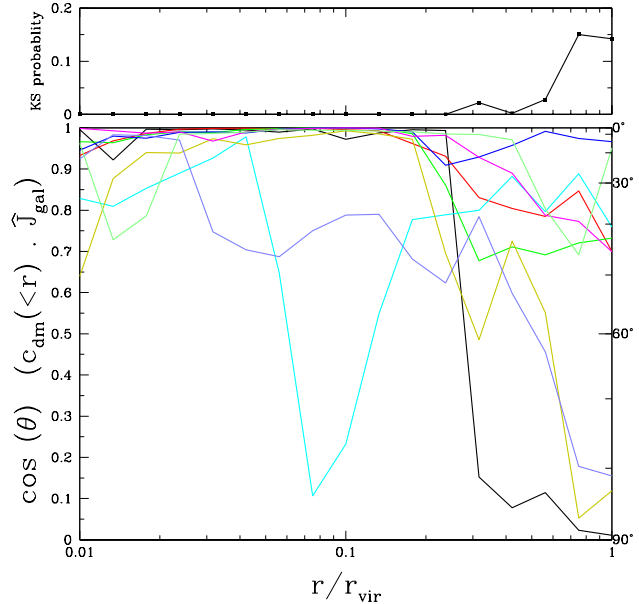} 
\caption{The angle
between the angular momentum of the disc ($\Jdisc$) and the short axis
of the dark matter halo ($\hat{\rm c}_{\rm dm}$) for the 9 galactic
system as a function of radius. Each of the 9 galactic halos is
plotted using a different colour, according to the same code used in
Fig.~8.}  
\label{Jdisccdm} 
\end{figure}

The orientation of the galaxy and its host halo is displayed in
Fig.~\ref{Jdisccdm} where we plot the angle between $\cdm$ and
$\Jdisc$ as a function of radius. Since $\Jdisc$ is always parallel to
$\hat{\rm c}_{\rm disc}$, this is equivalent to plotting the angle
between $\cdm$ and $\hat{\rm c}_{\rm disc}$. Apart, again, from the
anomalous gh9, the angular momentum vector of the galaxy points along
the short axis of the halo in the inner parts of the system. This
alignment begins to weaken beyond $\sim 0.5 r_{\rm vir}$, but even at
$r_{\rm vir}$ the KS probability is consistent with a uniform distribution
only at the $\sim 15$\% level. Our results agree well with those of
\cite{Bailin05} in the inner parts of the halo, but they are
marginally inconsistent at the virial radius where their small sample
of objects has a distribution of cosines consistent with uniform.

\begin{figure}
\includegraphics[width=20pc]{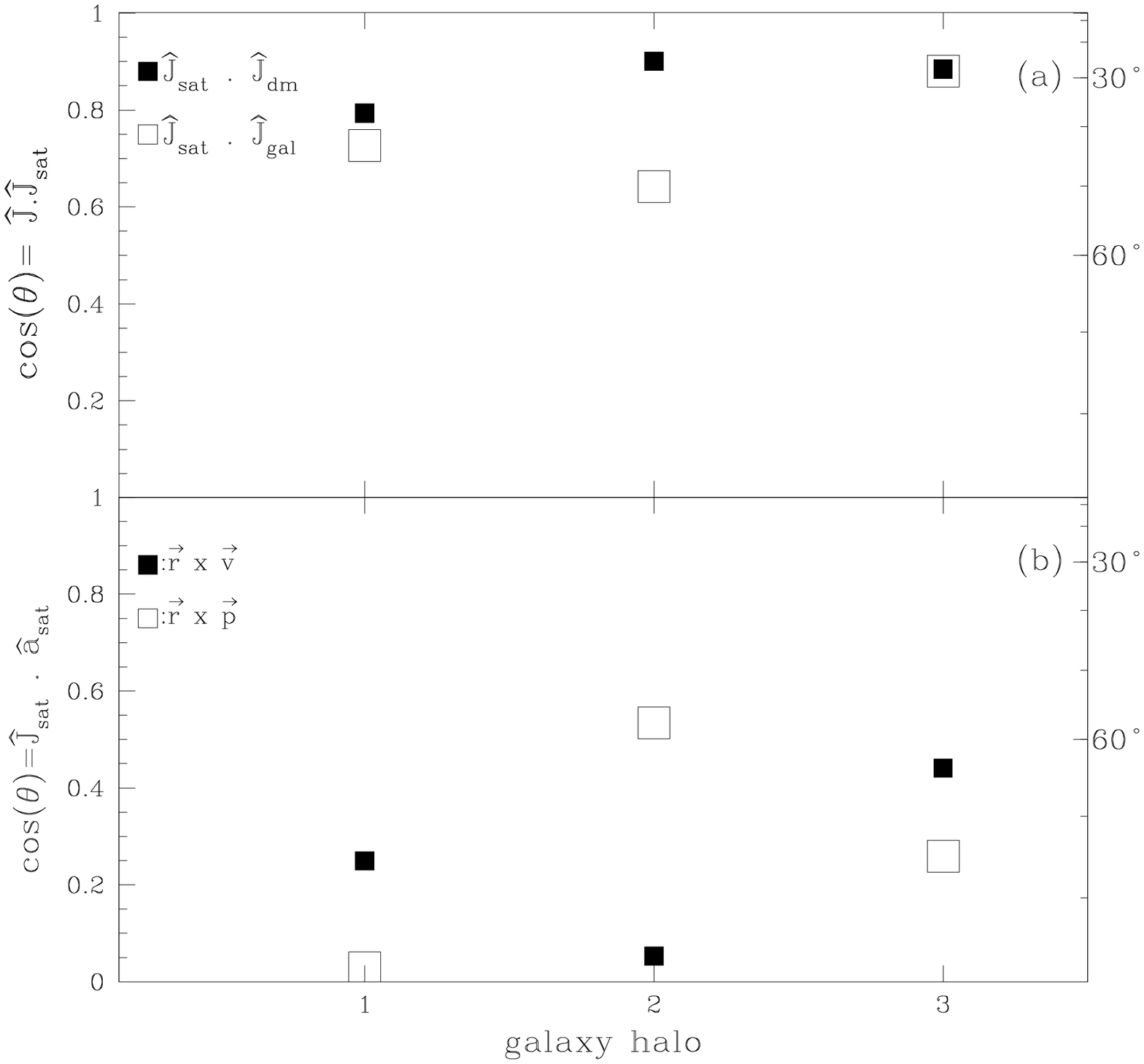}
\caption{\textit{Top panel} (a): the
cosine of the angle between the satellite angular momentum vector,
$\Jsat$, and $\Jdm$ (filled squares) and between $\Jsat$ and $\Jdisc$
(empty squares).  \textit{Bottom panel} (b): the cosine of the angle
between $\Jsat$ and the long axes of its spatial distribution, $\asat$. The
filled squares correspond to the case when the angular momentum is
calculated assuming each satellite has the same mass (i.e
$\underline{\rm J} = \langle \underline{r} \times \underline{v}
\rangle$), while the empty squares correspond to the case when the
angular momentum is calculated in the standard way (i.e
$\underline{\rm J} = \langle \underline{r} \times
\underline{p}\rangle$). }
\label{align.angmom}
\end{figure}

We now turn to the alignment of the angular momentum of the satellite
system with those of the galaxy and the dark matter halo. In this
case, we can use only the three systems with a large satellite
population. The results are plotted Fig.~\ref{align.angmom}(a) as
filled and empty squares.
For all three systems, there is some alignment (within $\sim
40^{\circ}$) between the spin vectors of the satellite systems and the
dark halo. However, in only one case (gh3) is the spin of the
satellites aligned with that of the galaxy. 

\begin{figure*}
\hbox{
\includegraphics[width=17.5pc]{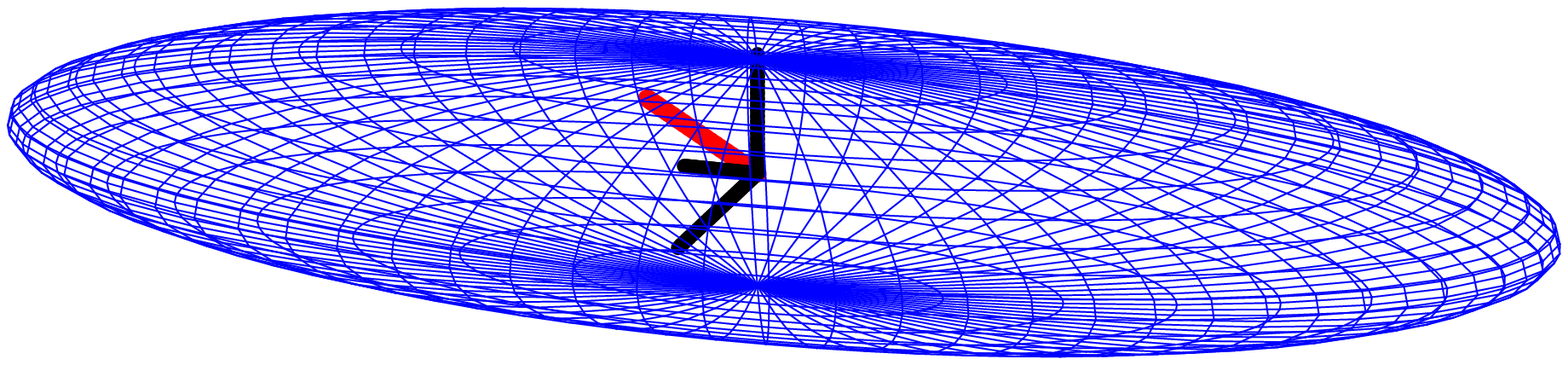}
\includegraphics[width=17.5pc]{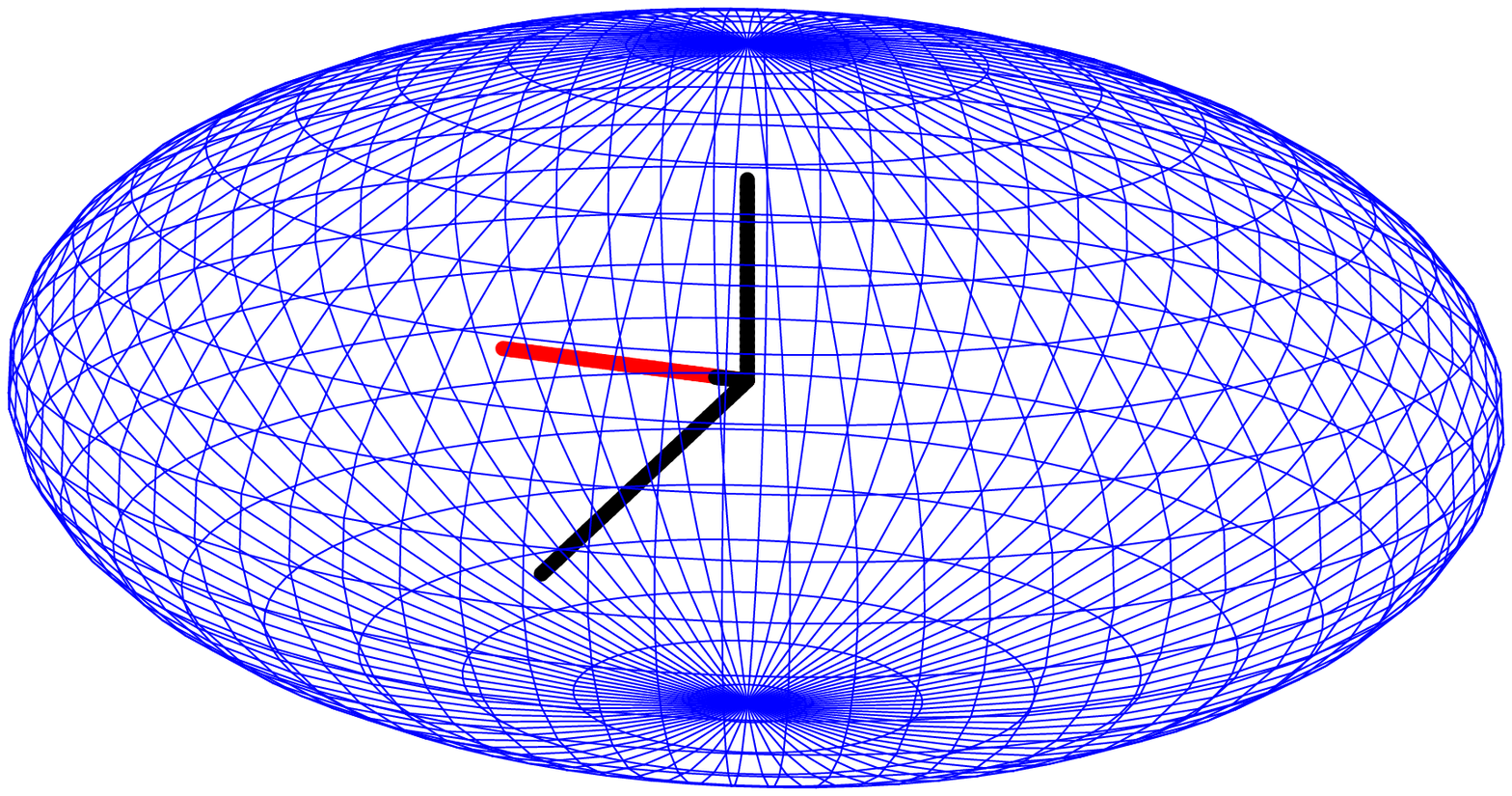}}
\includegraphics[width=17.5pc]{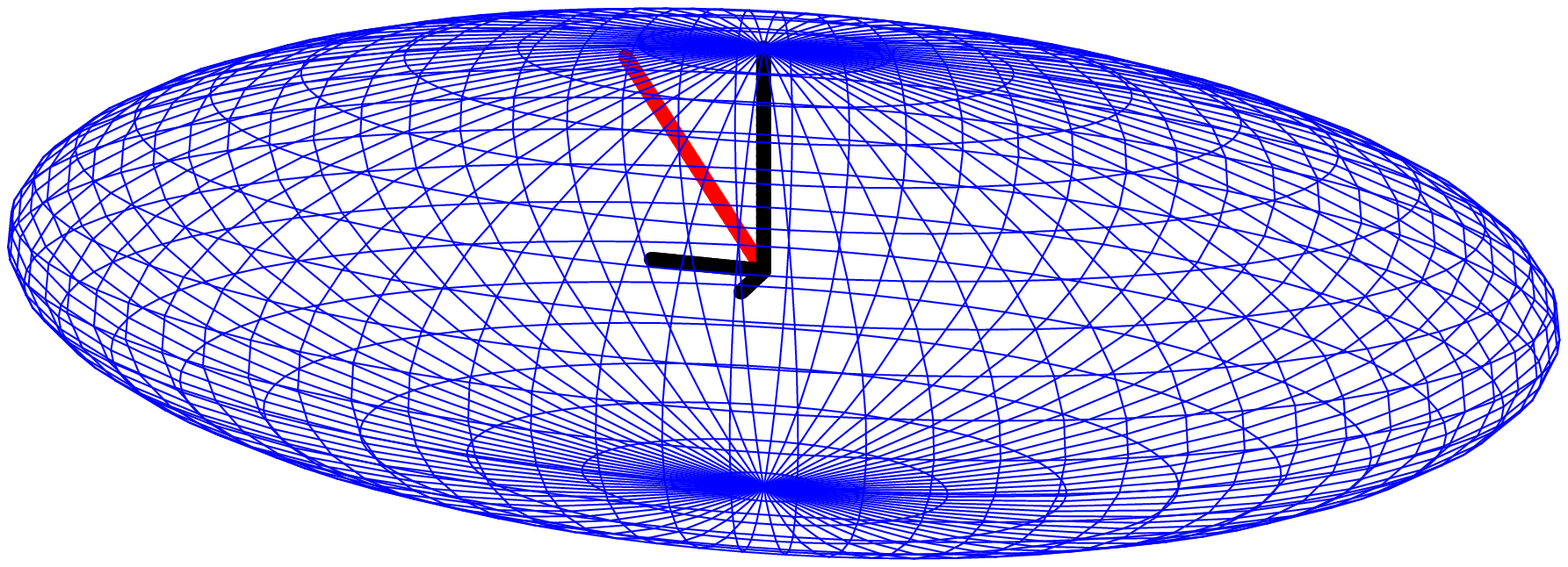}
\caption{Schematic diagram showing the satellite system angular
momentum vector, $\Jsat$, in three dimension (red line) calculated as
$\langle \underline{r} \times \underline{v} \rangle$. The blue
ellipsoids illustrate the shape of the satellite distribution. The
black lines are the projections of the vector $\Jsat$ along the three
principal axes of the satellite distribution. Clockwise, from the
left, we show galactic systems gh1, gh2 and gh3}
\label{3dorbits}
\end{figure*}

Finally, we ask whether the net angular momentum of the satellite
system is related to the shape of the system, i.e. whether the
satellites tend to orbit in the approximate plane that they define. If
this were the case, we would expect $\Jsat$ to point along the short
axis of their distribution, $\csat$. In fact, this is not what we
find. $\Jsat$ tends to point in a direction which, while often
approximately at right angles to $\asat$, the long axis of the
satellite system (Fig.~\ref{align.angmom}(b)), it is nevertheless not
along the short axis of the system, $\csat$. This arrangement is
illustrated schematically in Fig.~\ref{3dorbits} where the blue
ellipsoid demarks the structure defined by the anisotropic satellite
distribution, while the spin axis, $\Jsat$, is shown as a red line and
its three projections along the three principal axes of the satellite
structure are shown as black lines.
From Fig.~\ref{align.angmom}(b) we find that in all three cases the
angular momentum of the satellite system subtends an angle greater
than $\sim 60^{\circ}$ with the long axis of the satellite
distribution, indicating that the satellites are not orbiting in the
apparent plane defined by their spatial distribution. However, there
is little consistency amongst the 3 galactic systems which exhibit a
variety of orientations and alignments. 

\section{Conclusion and discussion}
\label{conclusion}

We have analyzed two N-body/SPH simulations of galaxy formation in a
$\Lambda$CDM universe, one of a single bright galaxy and the other of
a small region. Our simulations include the main physical processes
thought to be important in galaxy formation: metal-dependent gas
cooling, heating by photoionizing the primordial hydrogen gas, star
formation, metal production and feedback due to supernovae energy
injection. In total, we obtained a sample of 9 bright galaxies
comparable in luminosity to the Milky Way, containing a total of 78
satellites. Three of the central galaxies had a population of at least
11 satellites, the number of well-studied satellites in the Milky
Way. We have used these samples to investigate several interrelated
properties of the satellite population.

We first investigated the luminosity function of satellites, a
property often regarded as a challenge to the cold dark matter model
(e.g. \citealt{Moore99}). We find, however, that our simulations produce
an excellent match to the observed luminosity function of satellites
in the Local Group, at least to the resolution limit of the
calculation which corresponds to $M_V \sim -12$, close to the
luminosity of the faintest observed satellites.  The gas fractions in
these satellites agree well with observations, suggesting that their
formation paths may be similar to those of real satellites. Many of
the small subhalos resolved in the simulations fail to make a
substantially bright satellite. As \cite{Benson02b} have argued, star
formation is inefficient in small subhalos due to the combined effects
of reionization and supernovae feedback which limit the supply of cool
gas to the subhalo. 
 
The match of our simulations to the satellite data over most of the
observed range suggests that the relative paucity of satellites in the
Local Group (the ``satellite problem") does not reflect the much larger
abundance of subhalos, but rather feedback effects that limit the
growth of small galaxies. Furthermore, \cite{Benson02b} and
\cite{Stoehr02} have argued that the internal dynamics of CDM
satellites are also consistent with the data.  Overall, our results
agree well with the semi-analytic model of \cite{Benson02b}. In the
semi-analytic model, however, {\it bright} satellites, of luminosity
similar to the LMC or M33, were rare. This is not the case in
our simulations which match the bright end of the satellite
luminosity function well. The differences between the two treatments
are no doubt due to the different ways in which the physics of
galaxy formation are modelled, including differences in the treatment
of feedback and star formation and in the satellite merger rates.

We next considered the spatial distribution of the satellites. In the
outer half of the system, the number density profile of the satellites
tracks the dark matter well but in the inner parts it falls below
the dark matter profile. Remarkably, the number density profile of the
11 brightest satellites in one of our simulations is almost
indistinguishable from the profile for the 11 brightest satellites of
the Milky Way. The profiles of the other 2 well populated systems are
broadly similar, as is the profile for the satellites of M31.  As in
previous N-body studies (e.g. \citealt{Gao04}; \citealt{Libeskind06};
\citealt{Shaw06}), we find no resolved satellites in the inner $\sim
10\%$ of the halo.

We can study the distribution of individual satellite systems,
particularly their alignments with the halo and disc, for the three
systems that formed more than 11 satellites. In agreement with
previous purely N-body studies (\citealt{Libeskind05},
\citealt{Zentner05}), we find that the satellites tend to be
distributed in a highly flattened configuration whose major axis is
aligned with the major axis of the (generally triaxial) dark
halo. \cite{Libeskind05} argued that this arrangement reflects the
preferential infall of satellites along the spine of the filaments of
the cosmic web. Our gasdynamic simulations allow us to go further than
previous N-body work and investigate the alignment of satellite
systems with galactic discs and the alignment of the latter with the dark
matter halo. We find that in 2 out or 3 systems, the satellite system
is nearly perpendicular (to within $20^\circ$) to the plane of the
galactic disc. This surprising configuration is exactly what is seen
in the Milky Way. The third satellite system ended up well aligned
with the disc.

To investigate these alignments further, we calculated the
relationship between the disc and the halo, this time using the 9
bright galaxies in our simulations. Previous hydrodynamic simulations
have found a good alignment between the disc and the principal plane
of the halo (i.e. a good correlation between the directions of the
disc axis and the halo minor axis), but only in the inner $\sim 0.2
r_{\rm vir}$ of the halo (\citealt{Kazantzidis04};
\citealt{Bailin05}). Beyond this, these studies find little or no
correlation. We also find a good alignment in our simulations between
the disc and the halo at small radii but, unlike in the previous
studies, the correlation persists, albeit much weakened, out to the
virial radius. This sort of alignment is perhaps not unexpected in the
simplest interpretation of the tidal torque theory since both the dark
halo and the baryon component experience similar tidal torques.  The
relatively small differences between our results and those of
\cite{Kazantzidis04} and \cite{Bailin05} are most likely the result of
different treatments of star formation and feedback. For example, the
strong feedback in our simulations at early times which leads later to
a prolonged period of gas accretion and the formation of a large disc 
in a relatively quiet halo favours the persistence of a relationship
between the properties of the halo and the disc. 

Of the two cases in which the satellite systems are nearly
perpendicular to the disc, in the one which exhibits the strongest
alignment (gh1) the disc is also nearly perpendicular to the minor
axis of the halo. In the other case, the alignment between the disc
and the halo is weaker. In the third system (gh3), in which the
satellite system lies in the plane of the disc, the disc is only
roughly aligned with the halo. Although our sample is small, it
suggests that it should not be surprising that similar kinds of
alignments are found in observational studies of satellites in the
SDSS and 2dFGRS (\citealt{Brainerd05}, \citealt{Yang05}, \citealt{Sales04}).

Finally, we investigated the connection between the angular momenta of
the disc and the dark halo. For most systems, the two vectors are very
well aligned as a function of radius, out to the edge of the
system. The halo spin tends to be point along the short axis of the
halo. However, since the shape of the halo varies with redshift, the
spin axis of the disc or of the halo can often loose its alignment
with the short halo axis. Thus, care should be taken when interpreting
alignment statistics, either in simulations or in observational data.

In conclusion, our simulations have revealed a number of interesting
connections between the properties of central galaxies, their
satellite systems and their dark matter halos. They also point to a
good overall match between predictions of the $\Lambda$CDM model and
the observations of galaxy satellite systems that we have considered
here such as the luminosity function and the peculiar spatial
arrangement of the satellites of the Milky Way. However, due to their
high computational cost, our simulations are still too small to
provide good statistics. Larger simulations of this kind are required
to test the validity of the trends that we have found.

\end{document}